\documentclass[twocolumn,aps,prb,groupedaddress]{revtex4-2}

\usepackage{chngcntr}

\usepackage{graphicx}
\usepackage{color}
\usepackage{amsmath}
\usepackage{enumitem}
\usepackage{amssymb}
\usepackage{hyperref}
\usepackage{cancel}
\usepackage{ulem}
\usepackage{multirow}

\newcommand{\be}{\begin{equation}}
	\newcommand{\ee}{\end{equation}}

\newcommand{\bea}{\begin{eqnarray}}
	\newcommand{\eea}{\end{eqnarray}}

\newcommand{\lp}{\left(}
\newcommand{\rp}{\right)}

\renewcommand{\vec}[1]{{\boldsymbol #1}}

\begin{document}

\title{
Linear-in-temperature conductance in 
two-dimensional electron fluids}
\author{Serhii Kryhin, Qiantan Hong and Leonid Levitov}
\affiliation{Department of Physics, Massachusetts Institute of Technology, Cambridge, MA 02139}

\begin{abstract}
Linear temperature dependence of transport coefficients in metals is often ascribed to non-Fermi-liquid physics. Here we demonstrate the $T$-linear behavior of nonlocal conductivity in a clean 2D electron fluid, where carrier collisions assist conduction and lead to hydrodynamic transport with conductance rather than resistance growing with temperature. The key aspect is the occurrence of multiple hydrodynamic modes representing odd-parity modulations of the Fermi surface evolving in space and time. A cascade of such modes results in a linear $T$ dependence that extends to lowest temperatures, as well as a Kolmogorov-like fractional power $-5/3$ scaling of conductivity vs. wavenumber. These dependences provide a smoking gun for nonclassical hydrodynamics driven by such modes, expected to be generic for 2D electron fluids with simple near-circular Fermi surfaces.
\end{abstract}

\maketitle


Unusual temperature dependence of transport properties in metals 
often 
hints at exotic physics resulting from electron interactions. 
A celebrated example 
is the linear temperature ($T$) dependence of resistivity observed in many strongly correlated systems, particularly in the vicinity of quantum-critical points\cite{Sachdev2011,Phillips2022,Varma2020,Varma1989,Proust2019,Hartnoll2022,Zaanen2019,Hartnoll2015}.
Linear $T$ dependence of resistivity that extends to lowest temperatures has become a powerful tool for discerning non-Fermi-liquid phenomena that arise from strong interactions, such as scattering by emergent gauge fields and quantum-critical soft modes of different kinds\cite{Guo2022,Lee1989,Kim1994}. 
This stands in contrast with quadratic ($T^2$) scaling of transport coefficients predicted by conventional Fermi-liquid physics (see \cite{Pal2012,Lin_Behnia2015} and references therein).

Here our aim is to extend these ideas to the domain 
of electron hydrodynamics\cite{Gurzhi1968,Guerrero-Becerra2019,Hasdeo2021,Muller2009,Principi2016,Scaffidi2017, Narozhny2019,Alekseev2020,Toshio2020,Narozhny2021,Tomadin2014, Principi2016,Lucas2018,Qi2021,Cook2021,Valentinis2021a,Valentinis2021b,
Kiselev2019,tomogrph,Huang_Lucas2022,Hofmann2022,Hofmann2022b, Kryhin2021,
HGuo2017,AShytov2018,Nazaryan2021,Simon2020,Levchenko2020,Andreev2020,Abanov2020}. Recently it was emphasized that currents in electron fluids are governed by a nonlocal conductivity, featuring a power-law dependence on the wavenumber. Different exponent values have been identified for various regimes \cite{Kiselev2019,tomogrph,Huang_Lucas2022,Hofmann2022}, with the singularity at small wavenumbers indicating the significant role of long-wavelength modes. It is instructive to draw parallels to theories of non-Fermi-liquids, which predict fractional power-law scaling in frequency-dependent optical conductivity, such as $\sigma(\omega)\sim \omega^{-2/3}$ for $\omega\ll E_F$, and establish a connection between this fractional scaling and linear-in-$T$ resistivity 
\cite{Guo2022,Lee1989,Kim1994}. 

\begin{figure}[t]
\centering
\includegraphics[width=0.99\columnwidth]{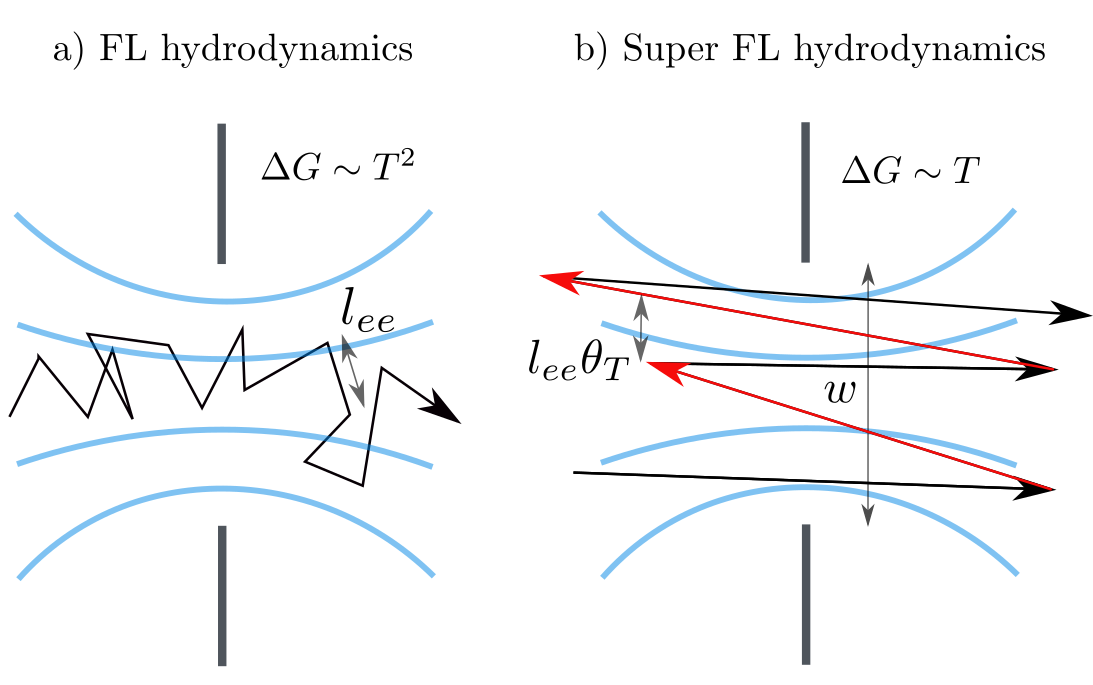} 
\caption{
a), b) Ordinary Fermi-liquid (FL) and unconventional super-Fermi-liquid hydrodynamic flows through a constriction. 
The FL phase occurs when 
the ee collision mean free path $l_{\rm ee}$ is smaller than the aperture. In super-FL phase the flow has a 
very different structure, manifested in a 
fractional-power divergence in $k$, Eq.\eqref{eq:G_vs_T}. 
At small $T$, typical trajectory 
features many backreflections at small angles $\theta_T\sim T/T_{\rm F}$, allowing particles to propagate out and then return as holes \cite{tomogrph,Kryhin2021}. 
Conductance behaves as $T^2$ for FL hydrodynamics and as $T$ for super-FL hydrodynamics, as detailed in Fig.\ref{fig3}. 
}
\label{fig1}
\end{figure}

In a similar vein, this paper will demonstrate a linear temperature scaling of nonlocal conductivity for two-dimensional electron fluids with simple Fermi surfaces: 
\be\label{eq:G_vs_T}
\sigma(k)\sim Tn k^{-5/3}
,
\ee
where $n$ is carrier density. The Kolmogorov-like fractional power-law dependence in Eq.\eqref{eq:G_vs_T} 
originates from a cascade of many long-lived modes. 
For a specific geometry of 
a constriction (Fig.\ref{fig1}) the quantity 
$\sigma(k)$ 
defines the conductance upon identifying the 
constriction width with $k^{-1}$.
This surprising behavior originates from nonclassical hydrodynamics\cite{tomogrph} driven by long-lived odd-parity excitations with super-Fermi-liquid  decay rates $\gamma_{\rm odd}\sim T^4$ which occur alongside the excitations with normal Fermi-liquid decay rates $\gamma_{\rm even}\sim T^2$ \cite{Kryhin2021, Hofmann2022}. 
These excitations create 
viscous modes with long lifetimes, 
giving rise to multi-mode electron hydrodynamics. 

Scaling linear in $T$ 
is clearly at odds with any analysis perturbative in microscopic rates behaving as $T^2$ and $T^4$. As we will see, Eq.\eqref{eq:G_vs_T} is 
a manifestly non-perturbative result, 
originating from a 
cascade encompassing a large family of coupled long-lived modes.
It is derived below for circular Fermi surfaces, ignoring momentum relaxation by disorder or phonons, however it 
holds for all non-circular Fermi surfaces that are convex and possess the $\vec p/-\vec p$ inversion symmetry, which insures the separation of time scales for the even and odd modes. Interestingly, the predicted nonclassical hydrodynamic response 
holds in a wide range of  $k$ extending to values 
beyond the nominal viscous/ballistic crossover $k\ell_{\rm ee}\sim 1$ (see Fig.\ref{fig3}).
 
 Transport measurements in graphene and GaAs constrictions 
 report on resistance decreasing with $T$ \cite{KrishnaKumar2017}
 and conductance growing with $T$\cite{Ginzburg2023}. In both measurements, the low-$T$ behavior of conductance, which exceeds ballistic values seen at $T=0$, is approximately linear in $T$. Superballistic conduction is a hallmark of hydrodynamic flow in which ee collisions facilitate conduction. Notably, a linear $T$ dependence that extends to lowest temperatures contradicts the conventional Fermi-liquid picture. 
This behavior provides 
a testable signature of the nonclassical hydrodynamics predicted here. 

\begin{figure}[t]
\centering
\includegraphics[width=0.99\columnwidth]{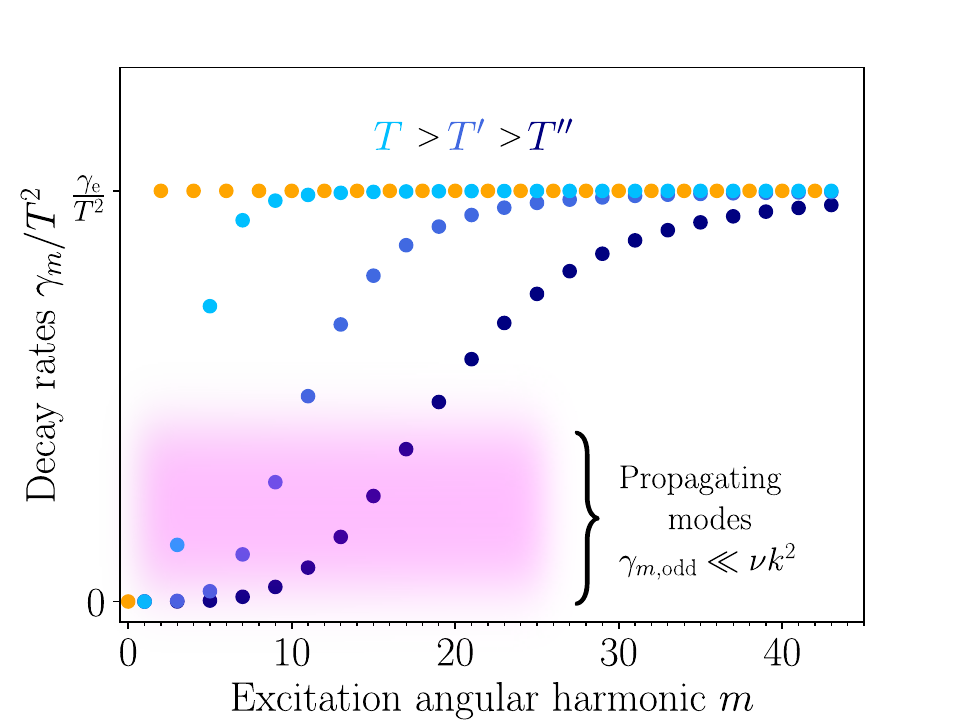} 
\caption{
Decay rates for even and odd excitations, $\gamma_m$,  vs. harmonic order $m$, 
shown as orange and blue dots, respectively \cite{tomogrph,Kryhin2021}. 
Pictured are rates for  
three different temperatures $T = 0.1 T_{\rm F}$, $T^\prime = 0.01 T_{\rm F}$, $T^{\prime \prime} = 0.003 T_{\rm F}$, with the lighter colors corresponding to higher temperatures.  Odd-$m$ rates scale as $T^4 m^4$ for small $T$ and not-too-large $m$. 
For large $m$ all rates scale as $T^2$ independent of the $m$ parity. For interpolation, we used the dependence 
$\gamma_{m\,{\rm odd}} = T^4 m^4 / (T_{\rm F}^2 + m^4 T^2)T_{\rm F}^2$. 
Highlighted in pink are modes with $\gamma_{m\,{\rm odd}} \lesssim \nu k^2$, which propagate as weakly damped viscous modes. The number of these modes grows rapidly as $T$ decreases, resulting in a 
cascade of soft modes and anomalous $T$ dependence of conductance. 
}
\label{fig2}
\end{figure}

The $T$-linear scaling contrasts the $T^2$ scaling expected for the conventional
Fermi-liquid  hydrodynamics governed by a single viscous mode\cite{Gurzhi1968}. In this case, the nonlocal conductivity, obtained from Stokes equation, is
\be\label{eq:sigma_Stokes}
\sigma(k)\sim \frac{n}{\nu k^2}
,
\ee
where $\nu=v_F^2/4\gamma_2$ is electrons' kinematic viscosity. For a Fermi liquid, $\nu$ scales as $1/T^2$, predicting a $T^2$ temperature dependence for $\sigma(k)$ and for transport through a constriction 
shown in Fig.\ref{fig1} \cite{HGuo2017}. 

In a realistic geometry such as that 
shown in Fig.\ref{fig1}, the conductance dependence on the aperture width 
$w$ and temperature can be inferred from the nonlocal conductivity $\sigma(k)$ by setting $k \approx 2\pi/w$. 
This allows to map out different regimes predicted for $\sigma(k)$ 
by tuning the 
lengthscale $w$. We illustrate the general relation between 
the $k$ dependence of $\sigma(k)$ and system lengthscale 
$w \approx 2\pi/k$ by the problem 
of transport through a long strip, for which a closed-form solution is available 
\cite{Kryhin2023}\cite{SI}.


The quantities that are central to our analysis of transport mediated by long-lived modes 
are the eigenvalues  $\gamma_m$ of the linearized collision operator of 2D electrons.  The values $\gamma_m$ give relaxation rates for different angular harmonics of the perturbed Fermi surface,\cite{Ledwith2019,tomogrph,Kryhin2021,Hofmann2022} 
\be
\delta f(\theta)\sim a_m \cos m\theta+b_m\sin m\theta
,
\ee
where $\theta$ is the angle on the Fermi surface. 
Odd-$m$ rates $\gamma_m$ are much smaller than even-$m$ rates and scale as $m^4$\cite{Ledwith2019,tomogrph}, forming a wide hierarchy of time scales
\be \label{eq:gammas_even_odd}
\gamma_{m\,{\rm even}}=\gamma
,\quad
\gamma_{m\,{\rm odd}}=\gamma' m^4
,\quad m \ll m_{\ast},
\ee
with the exception of $\gamma_0=\gamma_1=0$ for the density and velocity harmonics  $m=0$ and $1$, which are conserved.
At $T\ll \epsilon_F$, the rates depend on $T$ as 
\be
\gamma\sim T^2/\epsilon_F,\quad \gamma'\sim T^4/\epsilon_F^3
\ee
such  that $\gamma'\ll \gamma$ \cite{Kryhin2021,Hofmann2022}. Odd-$m$ rates $\gamma_m$ initially grow as $m^4$, saturating at the value $\gamma$ at a large 
\be\label{eq:m>m_*}
m \gtrsim m_{\ast}= (\gamma/\gamma')^{1/4}
\ee
This hierarchy of even-$m$ and odd-$m$ rates, pictured in Fig.\ref{fig2}, demonstrates that the family of long-lived modes, highlighted by pink in Fig.\ref{fig2}, quickly grows as temperature decreases. Because of their exceptionally low 
decay rates, these modes strongly contribute to hydrodynamics.

Our analysis will focus on
the nonlocal current-field response due to carrier movement governed by these modes. 
The response function of interest is nonlocal conductivity 
\be\label{eq:sigma_nonlocal}
j_\alpha(\vec x)=\int d^2x' \sigma_{\alpha\beta}(\vec x-\vec x')E_\beta(\vec x')
.
\ee
Expressed through the $k$-dependent conductivity as
\[
\sigma_{\alpha\beta}(k)\!\! =\int\!\! d^2 xe^{-i\vec k(\vec x-\vec x')} \sigma_{\alpha\beta}(\vec x-\vec x')
=\sigma(k)(\delta_{\alpha\beta}-\hat{\vec k}_{\alpha}\hat{\vec k}_{\beta})
\] 
it plays the same role in the theory of hydrodynamic response as Pippard's nonlocal $j$ vs. $A$ 
relation 
in the theory of superconductivity \cite{Tinkham} or the nonlocal current-field relation in the theory of the anomalous skin effect \cite{Pitaevskii}. However, unlike these 
effects, which describe the  distribution of supercurrents and normal currents in a narrow surface layer, here the relation in Eq.\eqref{eq:sigma_nonlocal} will be employed to describe current profile in the system bulk.

\begin{figure}[t]
\centering
\includegraphics[width=0.99\columnwidth]{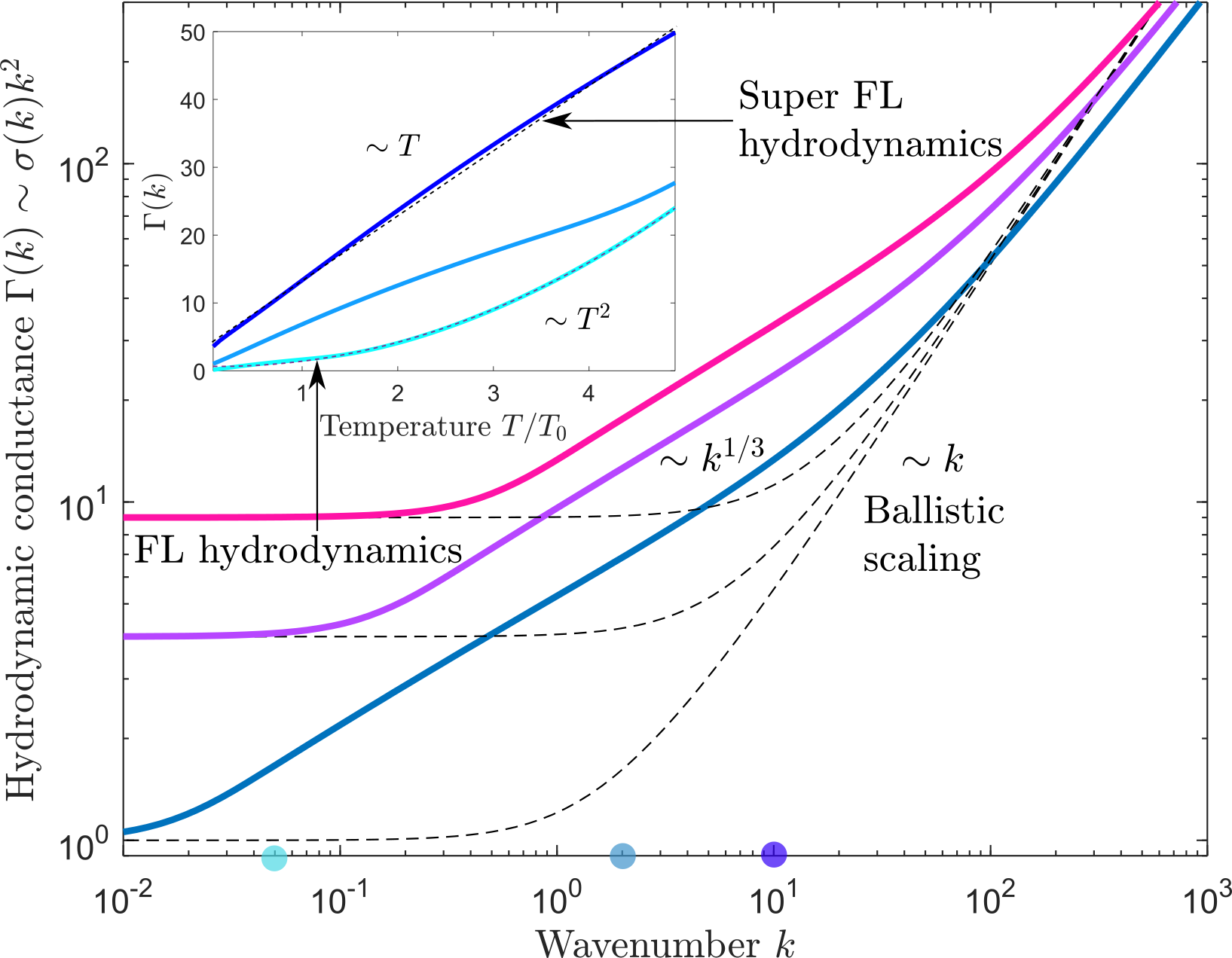} 
\caption{
Scale-dependent conductance $\Gamma(k)\sim \sigma(k) k^2$, Eq.\eqref{eq:sigma_transverse}, shown as a function of $k$ and $T$ 
for the rates $\gamma_m$ 
given in Fig. \ref{fig2}.
Main panel: $\Gamma(k)$ 
at three distinct temperatures $T_0$, $2T_0$, $4T_0$ (blue, purple and red curves) shown for $T_0 \approx 2.5 \cdot 10^{-3} T_{\rm F}$. The wavenumber $k$ is given in the units of $k_0=\gamma(T_0)/v_F$, obtained for $\gamma(T)$ taken at temperature $T_0$. The $k$ dependence 
features distinct regimes: Fermi-liquid (FL) 
and super FL hydrodynamic phases, in which $\Gamma(k)$ is $k$-independent and scales as $k^{1/3}$, respectively. 
As a comparison, $\Gamma(k)$ for conventional Gurzhi FL hydrodynamics is shown by dashed lines for matching parameter values (see discussion in text). 
Inset: $T$ dependence in these phases obtained for three $k$ values marked on the $x$ axis (these are taken in the FL, super FL and an intermediate phase---cyan, blue and light blue curves and dots). Temperature scaling, $T$ for super FL and $T^2$ for FL regimes, is validated by best fits to a linear and quadratic $T$ dependence (dashed lines, inset). 
}
\label{fig3}
\end{figure}

Fermion kinetics that accounts for perturbed Fermi surface dynamics yields a general closed-form solution for hydrodynamic conduction. This is achieved by linking 
the transverse conductivity $\sigma_{\alpha\beta}(k) =\sigma(k)(\delta_{\alpha\beta}-\hat{\vec k}_{\alpha}\hat{\vec k}_{\beta})$ to a quantity $\Gamma(k)$ defined as a continued fraction comprised of $\gamma_m$ rates \cite{Nazaryan2021,Kryhin2023} 
\begin{equation}\label{eq:sigma_transverse}
	\sigma(k) = \frac{D}{ \gamma_p+\frac{z}{\Gamma(k)}}
			,\quad
		    \Gamma(k)=\gamma_2+\frac{z}{\gamma_3+\frac{z}{\gamma_4+\frac{z}{\gamma_5+...}}}
		.
\end{equation}
Here $z=v_F^2 k^2/4$ and $D=ne^2/m$ is the Drude weight, where we expressed carrier density through Fermi momentum,  $n=g p_F^2/4 \pi \hbar^2$, with $g$ the spin/valley degeneracy.
The quantity $\Gamma(k)$ can be 
referred to as `hydrodynamic conductance'. 
Indeed, in the ideal fluid phase, i.e. in the absence of momentum relaxing scattering, $\gamma_p=0$, the conductivity $\sigma(k)$ equals $(D/z)\Gamma(k)$. 

The properties of $\Gamma(k)$ are sensitive to 
the even/odd parity asymmetry and $m$ dependence in $\gamma_m$. In Fig.\ref{fig2} we present numerical results for $\Gamma(k)$ obtained from the rates given in Eq.\eqref{eq:gammas_even_odd} 
The dependence $\Gamma(k)$ vs. $k$ exhibits three distinct regimes. 
Fermi-liquid hydrodynamics occurs at the smallest $k$. 
Here $\Gamma(k)$ is $k$-independent, giving conductivity scaling $\sigma(k)\sim 1/k^2$, as expected for linearized Navier-Stokes equation. In the `super-FL hydrodynamics'  phase 
$\Gamma(k)$ obeys power-law scaling $k^{1/3}$. Temperature dependence of $\Gamma(k)$, shown in the inset, is $T^2$ in the first regime, as expected for FL.  
Yet, a linear $T$ dependence emerges in the super-FL phase, a surprising behavior explained below 
(see Eqs.\eqref{eq:G_vs_T},\eqref{eq:sigma_Stokes}). At even higher $k$ a ballistic regime sets in, with $\Gamma(k)$ linear in $k$ and temperature-independent. Notably, the $T$-dependent part of $\Gamma(k)$, while being small at large $k$, remains linear, explaining the robustness of the $T$-linear scaling even at lowest temperatures. 

As evident in Fig.\ref{fig3}, the range of $k$ in which super-FL regime occurs grows as temperature decreases. 
This behavior 
can be understood as follows. 
At small $k$ the super-FL regime is bordered by a crossover to  Gurzhi's FL hydrodynamics. 
The crossover occurs at $k\approx \kappa_{<}$ corresponding to distances over which 
the longest-lived odd-$m$ excitation, namely the $m=3$ harmonic, travels over its lifetime. This gives a condition $\gamma_3=\nu \kappa_{<}^2$. 
On the large $k$ side it is bordered by a crossover from super FL to the ballistic regime. A simple estimate for this crossover is provided by a toy model \cite{Kryhin2023} in which  $\gamma_m$ for $m_\ast$ lowest odd-$m$ modes are neglected, whereas for $m>m_\ast$ the odd-$m$ and even-$m$ rates are taken to be equal. In this model the conductivity at small $k$ behaves as $\sigma_{\rm sFL}(k)=\frac{m_\ast+1}{\nu k^2}D$.  We set $m_\ast$ to the value at which the 
even/odd asymmetry  disappears, 
Eq.\eqref{eq:m>m_*}. The  crossover position $k_{>}$ is then determined by the condition $\sigma_{\rm sFL}(k)=\frac{D}{v_Fk}$. This gives the domain of existence of the super FL regime
\be\label{eq:super_FL_range}
\kappa_{<}<k<\kappa_{>}
,\quad \kappa_{<}=\frac{(\gamma\gamma')^{1/2}}{v_F}
,\quad \kappa_{>}=\frac{\gamma^{5/4}}{v_F{\gamma'}^{1/4}}
.
\ee
Since at low temperatures the rates $\gamma$ and $\gamma'$ scale 
as $T^2$ and $T^4$, 
the quantities $\kappa_{<}$ and $\kappa_{>}$ scale as $T^3$ 
and $T^{3/2}$, respectively. 
This explains widening of the super FL regime at low $T$. These estimates, as will be seen below, agree with the exact analytic result for $\sigma(k)$ scaling.

%
%



As a comparison, $\Gamma(k)$ for 
FL hydrodynamics is shown in Fig.\ref{fig3} 
(main panel, 
dashed lines). The $T$ dependence is absent in the ballistic regime and is $T^2$ in the FL regime. Notably, the $k$-linear behavior at large $k$, describing ballistic transport, is pushed to high $k$ in the super FL regime. This indicates that the super FL regime is readily accessible for realistic system widths.

Next, we present an analytic approach that allows to evaluate $\Gamma(k)$ and derive the $T$-linear and $k^{-5/3}$ scaling for super FL 
given in Eq.\eqref{eq:G_vs_T}. 
We first briefly consider an artificial case when all $\gamma_m$ 
take identical values $\gamma_m=\gamma$ ($m>1$) with no even/odd asymmetry. In this case evaluating the continued fraction yields $\sigma(k)$ of the form in Eq.\eqref{eq:sigma_transverse}  with $\Gamma(k)=( \gamma+\sqrt{\gamma^2+4z^2})/2$. 
In the long-wavelength limit $kv_F\ll\gamma$, the term $4z^2$ under the square root can be ignored, giving a result identical to that found from Stokes hydrodynamics, 
$
\sigma(k)=
D/\lp \gamma_p+z/\gamma\rp
$.
This quantity features a $T^2$ scaling at $\gamma_p=0$, Eq.\eqref{eq:sigma_Stokes}.

A very different behavior arises for 
the rates with an even/odd asymmetry. 
This problem is considerably more challenging, in particular because
odd-$m$ rates depend on $m$, with large differences between successive even and odd $m$ occurring at $m<m_*$, Eq.\eqref{eq:gammas_even_odd}. 
First, to capture this behavior in a simplified model, we set $\gamma_m=\gamma_{\rm e}$ for all even $m$ and $\gamma_{\rm o}$ for all odd $m$, ignoring the $m$ dependence of $\gamma_{\rm e}$ and $\gamma_{\rm o}$. 
In this case, $\Gamma(k)$ can be evaluated exactly: 
\be
\Gamma(k)=
\frac12\sqrt{\frac{\gamma_{\rm e}}{\gamma_{\rm o}}}
( \sqrt{\gamma_{\rm o}\gamma_{\rm e}}+\sqrt{\gamma_{\rm o}\gamma_{\rm e}+k^2v_F^2})
.
\ee 
This motivates introducing ``level-$m$'' partial continued fractions, defined as
\begin{equation}\label{eq:s0.5}
\Gamma_m(k)=\gamma_m+\frac{z }{\gamma_{m+1}+\frac{z }{\gamma_{m+2}+\cdots}}
.
\end{equation}
These quantities, evaluated similarly, are
\be 
\Gamma_m(k)=b_m ( \sqrt{\gamma_{\rm o}\gamma_{\rm e}}+\sqrt{\gamma_{\rm o}\gamma_{\rm e}+k^2v_F^2} )/2
,
\ee 
with $b_{m\,{\rm even}}=\sqrt{\gamma_{\rm e}/\gamma_{\rm o}}$ 
and $b_{m\,{\rm odd}}=\sqrt{\gamma_{\rm o}/\gamma_{\rm e}}$. 
When the even/odd difference 
is significant, $\gamma_{\rm e}\gg \gamma_{\rm o}$, the quantities $\Gamma_m$ are much larger for even $m$ than for odd $m$.


Below, we upgrade this picture in order to treat the quasi-continuum of weakly decaying modes highlighted in pink in Fig.\ref{fig2}. 
At low $T$, despite the even/odd beating effect, $\Gamma_m$ is slowly varying with $m$ for each individual parity of $m$. 
%
%
It is then natural to analyze the dependence 
$\Gamma_m$ vs. $m$ for a fixed parity. 
The result for $\Gamma(k)$ can then be found by taking $\Gamma_{m=2}$ \cite{SI}.
This analysis 
predicts an exact fractional-power-law scaling relation for $\Gamma(k)$:
\be\label{eq:Gamma_k_scaling}
  \Gamma(k)= 
  \frac{A\gamma}{(\gamma\gamma')^{\frac{1}{6}}}
  \lp\frac{kv_F}2\rp^\frac13,
	\quad
	A = \frac{\Gamma(\frac{1}{6})}{
	12^{\frac{2}{3}}\Gamma(\frac{5}{6})}\approx 0.94
	.
\ee
From the temperature dependence of the rates $\gamma\sim T^2$ and $\gamma'\sim T^4$, the resulting $T$ dependence of $\Gamma(k)$ is linear, 
\be
\Gamma(k)\sim T k^{1/3}.
\ee 
Eq.\eqref{eq:sigma_transverse} then yields 
$
\sigma(k) = 
		D/\lp \gamma_p+A'k^{5/3}/T\rp
$
with $A'$ a $T$-independent factor. The dependence on the wavenumber  $k$ translates into a dependence on system geometry. 
In the presence of momentum relaxing scattering due to phonons, in which case $\gamma_p\sim T$, the expected $T$ dependence becomes nonmonotonic, growing at low $T$, where ee scattering dominates, and decreasing at elevated $T$. 



To derive the scaling relation given in Eq.\eqref{eq:Gamma_k_scaling}, we perform
the recursion analysis of partial continued fractions, Eq.\eqref{eq:s0.5}, focusing on 
the even-$m$ quantities $\Gamma_m$.
Taking a difference of $\Gamma_m$ and $\Gamma_{m+2}$, and using Eq.\eqref{eq:s0.5} yields 
\begin{equation}
  \label{eq:s1}
  \Gamma_{m}-\Gamma_{m+2}=\gamma_m-\frac{\gamma_{m+1}\Gamma_{m+2}^2}{\gamma_{m+1}\Gamma_{m+2}+z }
\end{equation}
It turns out, perhaps surprisingly, that this exact nonlinear relation is greatly simplified 
through the substitution
\begin{eqnarray}\label{eq:s1.1}
\Gamma_m = \frac{z }{ \gamma_{m-1}}\left(\frac{u_{m}}{u_{m+2}}-1\right)
.
\end{eqnarray} 
After this substitution Eq.\eqref{eq:s1} yields linear relations
\begin{align}\nonumber 
\frac{z }{\gamma_{m+1}}u_{m+4}+\frac{z }{\gamma_{m-1}}u_m
=\left(\gamma_m+\frac{z }{\gamma_{m-1}}+\frac{z }{\gamma_{m+1}}\right)u_{m+2}
,
\end{align}
which represents a discrete second-order ODE for $u_m$. At low $T$, $\gamma_m$ and $u_m$ are slowly varying functions of $m$ in the region marked in pink in Fig.\ref{fig1}. 
The analysis based on slowly varying variables $u_m$ describes a cascade of modes
that propagate, slowly decay and interact with each other. In this case, the discrete ODE can be taken to the continuous domain, giving a Bessel equation that can be readily solved to obtain 
$\Gamma(k)$ given in Eq.\eqref{eq:Gamma_k_scaling} \cite{SI}.


Importantly, the transformation to a continuous domain is only valid when $|u^\prime(m)/u(m)| \ll 1$. It is therefore applicable in some of the transport regimes but fails in other regimes. For instance, the conventional hydrodynamic regime, which occurs at small wavenumbers $k$, is dominated by the $m=1$ velocity mode, whereas $u_{m>1}$ are negligible. It is therefore not described by $u_m$ which is quasi-continuous in $m$. Likewise, in the ballistic regime, which occurs at large enough $k$, the relevant harmonics encompass $m$ far outside the pink region in Fig.\ref{fig1} where $\gamma_{\rm odd}\ll \gamma_{\rm even}$. In contrast, the super-FL regime, 
describing the cascade of many long-lived modes, 
is captured well by the quasi-continuous approximation.

\begin{figure}[bt]
\includegraphics[width=\columnwidth]{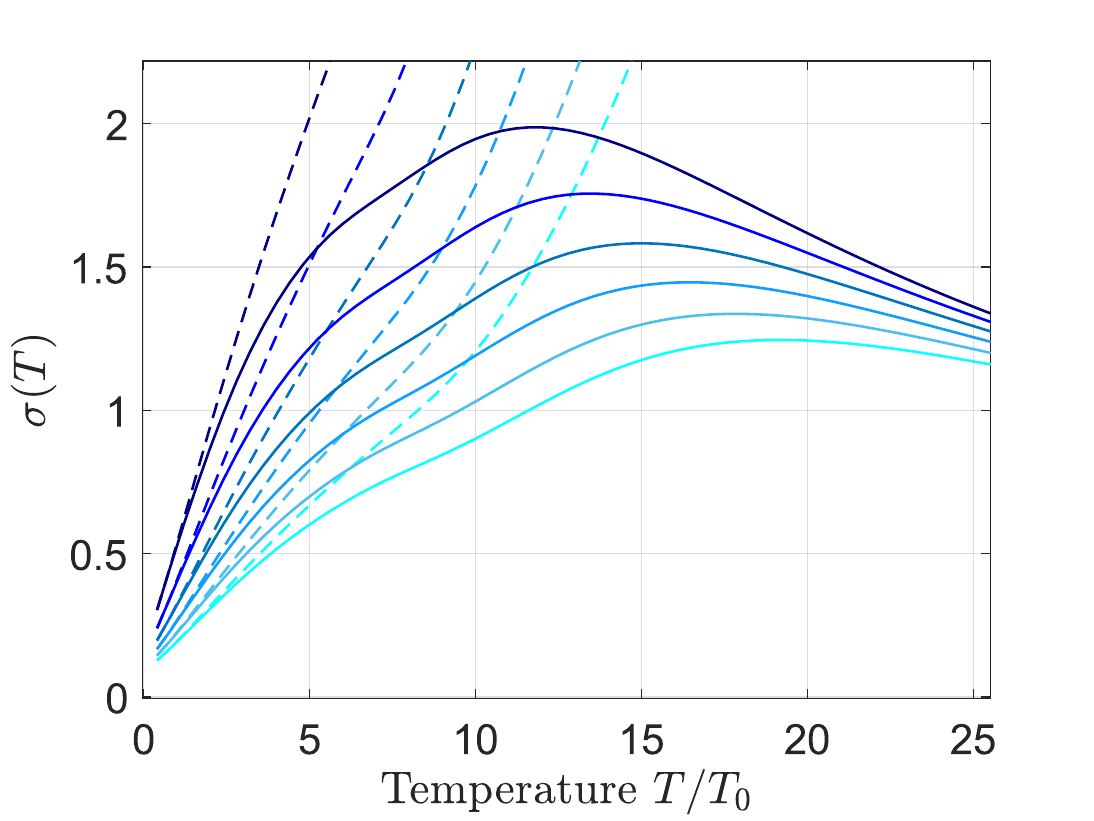} 
\caption{
Nonmonotonic $T$ dependence of 
conductivity arising in the presence of el-ph scattering [see Eqs.\eqref{eq:sigma_transverse},\eqref{eq:gamma_m_ee_elph}]. At low temperatures $T<T_{\rm BG}$ the el-ph scattering vanishes rapidly, giving negligible contribution to momentum relaxation, whereas at $T>T_{\rm BG}$ the momentum relaxation rate grows and dominates conductivity. Accordingly, the conductivity grows linearly with $T$ at low temperatures and decreases at higher temperatures (see text).
Shown are curves obtained for 
wavenumbers $k/k_0 = 10$, $12$, $14$, $16$, $18$ and $20$, where 
lighter colors correspond to smaller wavenumbers. The wavenumber and temperature units $k_0$ and $T_0$ 
are the same as in Fig.\ref{fig3}.
Dashed lines represent the conductivity in the absence of el-ph scattering, $\gamma_p = 0$. 
For illustration, the Bloch-Grüneisen temperature was taken to be $T_{\rm{BG}} = 6 T_0$. 
}
\label{fig5}
\end{figure}

Lastly, 
the knowledge of conductivity as a function of $k$ and $T$ can be used to predict transport coefficients 
for realistic geometry. As an illustration, we consider transport in a strip of width $w$ and length $\ell\gg w$. In this limit, we have an exact 
relation between the strip 
conductance $G=1/R$ 
and the nonlocal conductivity $\sigma(k)$ \cite{SI}: 
\be\label{eq:R_strip}
G=\frac1{\ell}\sum_{n=\pm1,\pm2...}\sigma(k_n) 
,\quad
k_n=2\pi n/w
.
\ee
As a sanity check, for 
$\sigma(k)=D/\nu k^2$ (the conventional viscous regime)
Eq.\eqref{eq:R_strip} predicts scaling 
expected for Poiseuille  flow, 
$G\sim DT^2 w^3/\nu L$. At $w\approx \ell$ this result agrees with expectations for viscous transport in constrictions \cite{HGuo2017}. 
In the super-FL regime described by 
$\sigma(k)\sim \frac{T}{k^{5/3}} n$, Eq.\eqref{eq:G_vs_T}, the strip conductance behaves as 
\be
 G \sim \frac{ T w^\frac{8}{3}}{ \ell}n
 .
\ee
At $L\approx w$ this yields $w^{5/3}$ scaling that agrees with the nonlocal conductivity scaling $k^{-5/3}$ derived above. 
It also matches the quantity $\sigma(k, T)$ with 
$k$ linked to the strip width as $k\sim w^{-1}$. Combined with the existence conditions for the super-FL regime in Eq.\eqref{eq:super_FL_range}, this defines a range of strip widths 
at which this regime can occur: 
\be
\frac{T^3}{T_{\rm F}^2} \lesssim \frac{\hbar v_F}{w} \lesssim T \sqrt{\frac{T}{T_{\rm F}}}.
\ee
The temperature interval in which the super-FL phase occurs rapidly widens as $T$ 
decreases. 


\begin{figure}[t]
\includegraphics[width=0.99\columnwidth]{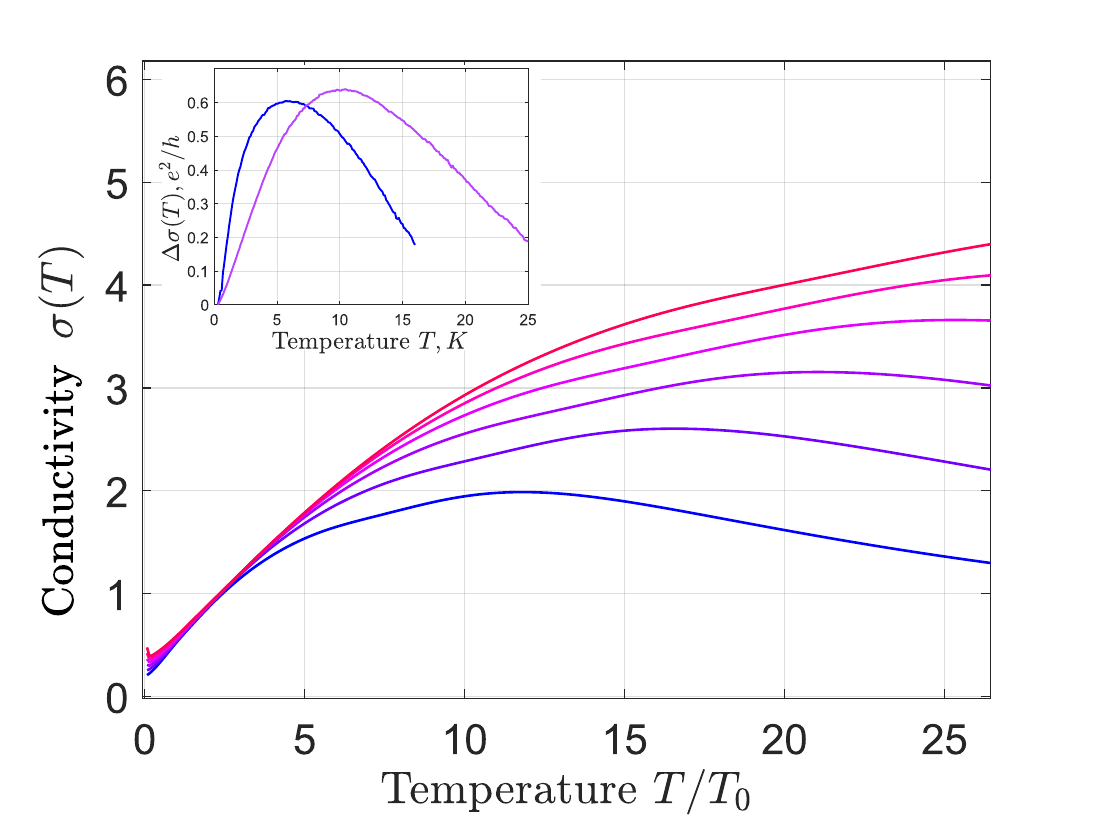} 
\caption{
Nonmonotonic $T$ dependence of conductivity, obtained as in Fig.\ref{fig5}, 
shown for several carrier density values: $n/n_0 = 1.0$, $1.5$, $2.0$, $2.5$, $3.0$, and $3.5$, where $n_0=(gm/2\pi\hbar^2)T_{\rm F}$ corresponds to the carrier density at the Fermi energy $T_{\rm F}$. 
The warmer colors correspond to higher densities. 
and a fixed $k/k_0 = 10$ was assumed.
The density dependence for the physical quantities 
used in  Eq.\eqref{eq:sigma_transverse} was taken to be $D \sim n$, $v_F \sim \sqrt{n}$, $k_{\rm F} \sim \sqrt{n}$, $T_{\rm F} \sim n$, and $T_{\rm BG} \sim \sqrt{n}$, which matches the dependence  $T_{\rm{BG}} \approx sk_{\rm F}$. 
Inset shows 
conductance temperature dependence reported in Ref.\cite{Ginzburg2023}, measured 
in constrictions formed in split-gated GaAs electron systems. The two curves represent the $T$-dependent part of conductance obtained for two different carrier density values \cite{footnote}.
}
\label{fig4}
\end{figure}

Next, we demonstrate that the $T$-linear behavior of $\sigma(k)$ at low temperatures remains robust in the presence of electron-phonon (el-ph) scattering. The reason for such robustness is that the rates of momentum-relaxing el-ph processes quickly diminish at low temperatures.  For 2D electron systems, microscopic analysis predicts 
relaxation rates vanishing as $\gamma_m \sim T^7 m^2$ at temperatures below the Bloch-Grüneisen temperature, $T<T_{\rm BG}\simeq sk_{\rm F}$, where $s$ is sound velocity \cite{Stormer1990}. 
At these temperatures, because of the $T^7$ dependence, the el-ph rates are subleading to the el-el rates which govern the super-FL hydrodynamics. However, the el-ph rates become significant at $T\gtrsim T_{\rm BG}$, scaling linearly with $T$ as temperature grows. As a result, at elevated $T$ 
the el-ph rates dominate the el-el rates, 
resulting in conductivity decreasing with temperature. This nonmonotonic $T$ dependence of conductivity, which is illustrated in Figs. \ref{fig4} and \ref{fig5}, can be described by our continued fraction model, Eq.\eqref{eq:sigma_transverse}, by incorporating the el-ph rates $\gamma_p$ in the rates $\gamma_{m}$ as
\be\label{eq:gamma_m_ee_elph}
\gamma_m\to\gamma_m+\gamma_p
.
\ee 
Following Refs.\cite{Stormer1990}, 
we model the temperature dependence of 
$\gamma_p$ by an expression 
\begin{equation}\label{eq:el_ph_rate}
	\gamma_p \sim \left( \frac{T}{T_{\rm BG}} \right)^7 \int_0^{T_{\rm BG}/T} dt \frac{t^7}{(e^t-1)(1-e^{-t})},
\end{equation}
valid at $T$ both larger and smaller than $T_{\rm BG}$ (but excluding the $m^2$ dependence in the low temperature regime for simplicity). At such temperatures, $\gamma_p$ behaves at $T$ and $T^7/T_{\rm BG}^6$, respectively. Incorporating $\gamma_p$ 
into $\gamma_m$ 
impacts conductivity as illustrated in Figs.\ref{fig4} and \ref{fig5}: the linear-in-$T$ conductivity growth at low $T$ remains robust, with the linear behavior extending to lowest $T$. In contrast, at higher $T$, the sign of conductivity $T$ dependence is reversed, giving a nonmonotonic behavior which agrees with experimental observations reported in Refs.\cite{KrishnaKumar2017,Ginzburg2023}

In summary, conductivity growing  with temperature is a hallmark of the hydrodynamic regime in which ee collisions facilitate conduction and suppress dissipation. In contrast to the conventional FL hydrodynamics, which predicts conductivity growing as $T^2$ at $T\ll T_{\rm F}$, the super-FL conductance grows more rapidly, increasing linearly with temperature. The enhanced conductance reflects presence of multiple long-lived modes unique to 2D fermions. Recently, conductance exceeding the ballistic $T=0$ value and growing linearly with $T$ has been observed for superballistic transport in graphene and GaAs electron fluids \cite{KrishnaKumar2017,Ginzburg2023}. 
This indicates that ongoing experiments 
\cite{Sulpizio2019,Ku2020,Braem2018,Vool2021,Aharon-Steinberg2022}
are well positioned to probe the exotic hydrodynamics driven by emergent long-lived modes. 
Further, we expect 
similar anomalous $T$-dependent hydrodynamic effects in strongly-correlated systems that host soft quantum-critical modes or emergent gauge fields \cite{Sachdev2011,Phillips2022,Varma2020,Varma1989,Proust2019,Hartnoll2022,
Zaanen2019,Hartnoll2015, 
Guo2022,Lee1989,Kim1994} activated at low temperatures.  

We are grateful to Klaus Ensslin, Andre Geim, Lev Ginzburg and Boris Spivak for useful discussions. This work was supported by the Science and Technology Center for Integrated Quantum Materials, National Science Foundation grant No.\,DMR1231319 
and was performed in part at Aspen Center for Physics, which is supported by 
NSF grant PHY-2210452. 
SK is currently affiliated with Harvard University, Physics Department. 


\bigskip
\newpage

\centerline{\large{\bf Supplemental information}}
\centerline{\large for ``Linear-in-temperature conductance in 
}
\centerline{\large two-dimensional electron fluids''}
\centerline{\large by S. Kryhin, Q. Hong and L. Levitov}

\section{Geometry-dependent conductance} 

Here we detail properties of the geometry-dependent conductance referred to 
in the main text. Namely, 
for a realistic system geometry studied in Refs.\cite{tomogrph,Kryhin2023}, the dependence of the conductance 
on system width and temperature can be directly inferred from the $k$ and $T$ dependence of the nonlocal conductivity $\sigma(k,T)$. 
 Specifically, we 
demonstrate the implications of 
the nonlocal conductivity  scaling with $k$ and $T$. 

The geometry frequently used in transport measurements to probe hydrodynamic behavior is that of a constriction. We model it as a narrow channel of width $w$ and length $\ell$. Using a closed-form analytic solution available for a long channel \cite{tomogrph,Kryhin2023}, $\ell\gg w$, we show that:
\begin{enumerate} 
\item The system conductance is a sum of contributions $\sigma(k)$ with wavenumbers $k\gtrsim 2\pi/w$, where the dominant contribution originates from $k\approx 2\pi/w$.
\item Therefore, the conductance dependence on the system width $w$ and temperature is well described by the dependence $\sigma(k,T)$  
for $k\approx 2\pi/w$. 
\end{enumerate}
The behavior of the channel conductance can be predicted by combining these properties with the power-law dependence $\sigma(k)\sim T k^{-5/3}$ for the nonlocal conductivity. 
This scaling relation, derived in the main text for the super FL phase for wavenumbers $\kappa_{<}<k<\kappa_{>}$, will 
translate into a scaling of the channel conductance 
%
\be
G(w)\sim T w^{8/3}/\ell
,
\ee
inferred at $k\approx 2\pi/w$. 

As discussed in the main text, the $\sigma(k)$ scaling holds for $\kappa_{<}<k<\kappa_{>}$, where $\kappa_{<}$ and $\kappa_{>}$ describe  crossovers to the FL hydrodynamic and ballistic phases, respectively. Since at low temperatures $T\ll T_{\rm F}$ the crossover positions behave as $\kappa_{<}\sim T^3$ and $\kappa_{>}\sim T^{3/2}$, the range of $k$'s in which super FL hydrodynamics occurs quickly widens as $T$ decreases. Furthermore, for a channel of width $w$ the condition $\kappa_{<}<k$ is always fulfilled at low enough $T$. As a result, the $T$-linear behavior dominates at low temperatures. Indeed, as discussed in the main text, the $T$-linear scaling is characteristic of both the super-FL phase and the ballistic phase. It is a dominant contribution for the former and 
represents a residual $T$-dependence on top of a $T=0$ value for the latter. Therefore, the Fermi-liquid $T^2$ conductivity scaling is always superseded by a $T$-linear scaling at low enough temperatures. Once the $T$-linear dependence sets in, it extends all the way to the lowest temperatures, first through the super FL phase and then, at lower $T$, through the ballistic phase. This justifies the conclusion that the linear-in-$T$ behavior is robust at all low temperatures, in agreement with the numerical results presented in the main text. 

The inequalities $\kappa_< < k < \kappa_>$ can be used to estimate the device length at which one can observe the $T$-linear scaling. As discussed above, the order of magnitude for $k$ in realistic geometries is set by the device size $w$ and results in $k \approx 2 \pi/w$. In the main text we provide the relations between the values of $\kappa_<$ and $\kappa_>$ including all the dimensional prefactors. Eqs. 4 and 8 of the main text lead to the crossover between the ballistic and super-FL regime occurring at the system sizes of the order of
\begin{equation}
	w' \sim \frac{\hbar v_F}{\varepsilon_F} \left( \frac{T_{\rm F}}{T} \right)^{3/2}.
\end{equation}
The crossover between super-FL regime and  ordinary hydrodynamics happens at
\begin{equation}
	w'' \sim \frac{\hbar v_F}{\varepsilon_F} \left( \frac{T_{\rm F}}{T} \right)^3.
\end{equation}
The same pair of conditions on $w$ is obtained with a use of different approach in the main text (see Eq. 16).
Thus for $T \ll T_{\rm F}$ there is a large window of device lengths where the super-FL regime takes place. For conventional FL hydrodynamics, the onset length scale is the mean free path for 2-body collisions $v_F/\gamma$, where gamma is the standard Landau Fermi liquid rate $\gamma \sim T^2/\hbar \epsilon_F$. In a realistic system this length scale must be smaller than the aperture width and disorder free path, etc. In our problem, however, because of the $T^{3/2}$ temperature dependence predicted for the ballistic/super-FL crossover, the onset of hydrodynamics is reachable more easily than for the conventional hydrodynamics (owing to $3/2<2$).



Moving on to the analysis of the transport problem, 
we first describe 
a framework \cite{tomogrph,Kryhin2023} that links the scale-dependent conductivity $\sigma(k)$ 
to the linear geometry-dependent conductance response of the system in direct space.  
We first consider the method in a most general form and then specialize to 
the strip geometry, assuming for now that the length is much greater than the width, $\ell\gg w$, and taking the limit $\ell\approx w$ at the end. 

The most general current-field linear response due to a space- and time-dependent external electric field $E_i(\vec x, t)$ can be written as
\begin{equation}\label{eq:j_E_general}
	j_\alpha (\vec x, t) = \int d^3\vec x^{\prime} \, \sigma_{\alpha \beta}(\vec x, \vec x^\prime) E_\beta(\vec x^\prime, t),
\end{equation}
where indices $\alpha$ and $\beta$ 
label spatial vector components. The kernel $\sigma_{\alpha \beta}(\vec x, \vec x^\prime)$ is the non-local conductivity in real space. In the case of an infinite system, the response form is constrained by space translation and rotation symmetries as  
$\sigma_{\alpha \beta}(\vec x, \vec x^\prime) = \sigma_{ij}(\vec x - \vec x^\prime)$. 

In a system of a finite size, finding the current-field response requires solving a boundary-value problem posed for the transport equations for carrier distribution with suitable boundary conditions. In general, this yields conductance that may have a nontrivial dependence on system geometry. Since this problem is well known to be mathematically challenging even when the constituting current-field response relations are local (ohms' law), and can become considerably more cumbersome for the general nonlocal response as in Eq.\eqref{eq:j_E_general}, it is natural to seek alternative ways. 
Here, we employ the phenomenological method developed in \cite{tomogrph,Kryhin2023}, which allows to account for the boundary conditions for current $j_\alpha$ for a general sample geometry. The unique advantage of this method is that it utilizes the infinite-space translation-invariant conductivity to describe the geometry-dependent currents for a given system geometry.


Specifically, we focus on the conductance of a long strip 
of width $w$. We take the strip to lie in the $x-z$ plane and directed along the $z$ direction, 
with the boundaries at $x = 0$ and $x = w$. The electric field that drives the current is directed along the $z$-axis. The current and field in the strip obey the nonlocal relation, Eq.\eqref{eq:j_E_general}. To solve this problem, instead of the strip geometry in question, 
we consider an auxiliary problem in an infinite plane. Namely, the strip problem is turned into a periodic problem in $x$ by positioning extra ``boundaries" at $x =wn$ for integer $n$. Further, we model the drag produced by the diffuse boundaries of the sample by introducing an internal electric field $\delta E\parallel z$ that is concentrated in the vicinity of the boundaries and is proportional to the current that runs in the vicinity of those boundaries. In the infinite-plane representation, we write the internal field localized at the boundary as
\begin{equation}\label{eq:boundary_field}
	\delta E(x) = - \alpha \sum_{p  = - \infty}^\infty j(x) \delta(x - wp).
\end{equation}
The coefficient $\alpha$ represents a phenomenological boundary drag effect that mimics carrier momentum relaxation due to dissipation caused by the diffuse scattering at the boundary. 

With this $\delta E(x)$ we can seek for the resulting current by self-consistently solving the conductance equation
\begin{equation}\label{eq:j_E_selconsistent}
	j_z(x) = \int d x^\prime \, \sigma_{zz} (x - x^\prime) \left( E_0 + \delta E(x^\prime) \right),
\end{equation}
where $E_0$ is an external homogeneous field applied along $z$-axis. The non-local conductivity $\sigma_{zz}(x - x^\prime) = \int \frac{dk_x}{2 \pi} e^{i k_x (x - x^\prime)} \sigma(k_x)$ defined in Eq.(6) of the main text, since in our setup $\vec k$ is directed along $x$-axis and is perpendicular to the direction of $E_0$ and $\delta E(x)$. Note that if we take the limit of $\alpha \to \infty$, in order for the response to remain finite the current on the boundaries should approach $0$. Therefore, the limit $\alpha \to \infty$ reproduces correctly the diffuse boundary conditions for the current $j$. 

The self-consistent solution of the problem in Eqs. \eqref{eq:boundary_field} and \eqref{eq:j_E_selconsistent}, after taking the limit $\alpha\to \infty$ and accounting for the periodicity in $x$, takes the form
\begin{align}\label{eq:jx}
	j(x) = j_0 \left( 1 - \frac{\sum_{n \neq 0} \sigma(k_n) e^{i k_n x}}{\sum_{n \neq 0} \sigma(k_n)} \right),
	\quad 
	k_n = \frac{2\pi n}{w},
	\end{align}
	where $n$ is an integer and 
\begin{align}	\label{eq:j0}
	j_0 = \sigma_\mathrm{eff}E_0, \quad \sigma_\mathrm{eff} = \sum_{n \neq 0} \sigma(k_n).
\end{align}
Note that since $j_0 w$ is the net current flowing through the strip, the quantity $\sigma_\mathrm{eff} w$ defines the conductance per unit length of the strip. Therefore, the geometry-dependent conductance of the strip is $\sigma_\mathrm{eff}w/\ell$. 

This result allows to illustrate the relation between the nonlocal conductivity $\sigma(k)$ and geometry-dependent conductance, described above (properties 1 and 2). Indeed, the quantity $\sigma_\mathrm{eff}$ is a sum of contributions $\sigma(k)$ for wavenumbers $k_n = 2\pi n/w$, all of which exceed the inverse strip width $2\pi/w$. Since $\sigma(k)$ is a monotonic function of $k$ that quickly drops to zero at large $k$, the sum giving $\sigma_\mathrm{eff}$ is dominated by the lowest-$k$ contribution, $\sigma_\mathrm{eff} \approx 2\sigma(k=2\pi/w)$. These properties agree with the behavior described above. 
Therefore, as discussed in the main text, the form of $\sigma(k)$ completely defines the scaling of the strip resistance with the carrier temperature $T$ and the strip width $w$.

\section{The cascade of long-lived modes and conductivity scaling} 

Here we detail the derivation of the scaling behavior of the nonlocal conductivity given in the main text, Eqs. (1) and (10). Our analysis relies on a continued fraction representation of conductivity. 
We introduced level-$m$  
continued fractions $\Gamma_m$ which obey recurrence relations, and used them to calculate 
$\Gamma_{m = 2}$ -- the quantity of interest.
To analyze the recurrence relations obeyed by $\Gamma_m$, we split them into  separate recurrence relations for $m$ of a fixed parity, even or odd, as follows: 
\be\label{eq:app_odd_even}
	 \Gamma_{m}-\Gamma_{m+2}=\gamma_m-\frac{\gamma_{m+1}\Gamma_{m+2}^2}{\gamma_{m+1}\Gamma_{m+2}+z }.
\ee
It was found convenient to perform variable substitution 
\be\label{eq:sub}
	\Gamma_m = \frac{z }{ \gamma_{m-1}}\left(\frac{u_{m}}{u_{m+2}}-1\right),
\ee
after which the recurrence relations greatly simplify.
Namely, this substitution 
yields a linear relation
\begin{align}\nonumber 
\frac{z }{\gamma_{m+1}}u_{m+4}+\frac{z }{\gamma_{m-1}}u_m
=\left(\gamma_m+\frac{z }{\gamma_{m-1}}+\frac{z }{\gamma_{m+1}}\right)u_{m+2}
. 
\end{align}
By regrouping terms, the relation above can be cast into the form 
resembling a discretized second-order ODE: 
\begin{align}\label{eq:s1.3}
&  \frac{1}{2}\left(\frac{z }{\gamma_{m-1}}+\frac{z }{\gamma_{m+1}}\right)(u_{m+4}-2u_{m+2}+u_{m})
\\ \notag
& -\frac{1}{2}\left(\frac{z }{\gamma_{m-1}}-\frac{z }{\gamma_{m+1}}\right)(u_{m+4}-u_{m})=\gamma_{m}u_{m+2}
.
\end{align}
We emphasize that these relations are totally general. 
Indeed, our starting point is a tridiagonal system of equations for the amplitudes of different harmonics\cite{Kryhin2023}. In this case there is a natural bipartite structure (the off-diagonal couplings are between harmonics of different parity). Eliminating variables of one parity yields a tridiagonal problem describing the other parity. 
Hereafter we take $m$ values 
to be even. 


Since $\Gamma_m$ and $\gamma_m$, when restricted to a fixed parity, are slowly varying with $m$, we take Eq.\eqref{eq:s1.3} to a continuous domain by replacing the differences with derivatives 
\begin{eqnarray}
4 \frac{z }{\gamma_{\rm o}} \frac{d^2u_m}{dm^2}-4 z  \frac{d \gamma_{\rm o}^{-1}}{dm} \frac{du_m}{dm}=\gamma_{\rm e} u_m \notag
,
\end{eqnarray}
where we introduced notation $\gamma_{\rm e}(m) = \gamma_m$ and $\gamma_{\rm o}(m) = \gamma_{m+1}$. This gives a second-order differential equation 
\begin{eqnarray}\label{eq:s2}
u_m''-\frac{\gamma_{\rm o}'}{\gamma_{\rm o}}u_m'-\frac{\gamma_{\rm o}\gamma_{\rm e}}{4z }u_m=0
,
\end{eqnarray}
where derivatives are taken with respect to $m$ treated as a continuous variable. 

In such continuous-domain representation the expression for $\Gamma_m$, Eq.\eqref{eq:sub}, reads
\begin{eqnarray}\label{eq:s3}
\Gamma(m) = -\frac{2z }{ \gamma_{\rm o} u_m}\frac{du_m}{dm}
.
\end{eqnarray}
Below, after solving the differential equation for $u_m$, we will use this relation to obtain $\Gamma(m)$.

Note that, while Eq. \eqref{eq:s1.3} is completely general, the continuous-domain ODE in Eq. \eqref{eq:s2} is only valid provided $|u^\prime / u | \ll 1$. This condition, which was introduced when replacing the discrete derivatives with continuous derivatives, is somewhat subtle. Indeed, it is valid in some of the transport regimes of interest but 
	inapplicable in other regimes. For example, when applied to the case of identical $\gamma_m = \gamma$ ($m>1$) it correctly reproduces the ballistic regime; yet, it does not  reproduce the ordinary hydrodynamic regime, since the condition $|u^\prime/u| \ll 1$ does not hold in this case. However, the continuous-domain representation and the condition $|u^\prime / u | \ll 1$ work well for super-FL hydrodynamics. This reflects the large number of coupled long-lived modes $0<m<m_\ast$, $m_\ast\gg 1$, which form the cascade that governs transport in this regime. 

Returning to the current model,
in which $\gamma_{\rm o}$ and $\gamma_{\rm e}$ are unequal and depend on $m$  as described in Eqs.(4) and (5) of the main text, 
\be\label{eq:m4_model}
\gamma_{\rm o} = \gamma' m^4,\quad \gamma_{\rm e} = \gamma,
\ee
we proceed as follows. 
The rates $\gamma_{\rm o}$ and $\gamma_{\rm e}$ converge to a constant value $\gamma$ for high harmonics, $m\gtrsim m_{\ast}=(\gamma/\gamma')^{1/4}$. For such $m$ the values $\Gamma(m)$ 
are given by the solution of the one-rate model $\Gamma(m)=(\gamma+\sqrt{\gamma^2+k^2v_F^2})/2$. Therefore, we can set the boundary condition for Eq.\eqref{eq:s2} as
\begin{equation}\label{eq:s1.5}
 \left.-\frac{2z }{ \gamma u}\frac{du}{dm}\right|_{m=m_{\ast}}=\Gamma(m_{\ast})=\frac{\gamma+\sqrt{\gamma^2+k^2v_F^2}}{2}
 ,
\end{equation}
and proceed to solve Eq.\eqref{eq:s2} on the interval $0<m<m_{\ast}$ to obtain the quantity of interest $\Gamma_{m=2}$.

Continuing to work with the $m^4$ model, Eq.\eqref{eq:m4_model}, 
we write Eq.\eqref{eq:s2} as
\begin{eqnarray}\label{eq:s3.5}
u_m''-\frac{4}{m}u_m'-\frac{\sqrt{\gamma\gamma'}}{4z }m^4u_m=0
.
\end{eqnarray}
To tackle this equation, we 
introduce a new variable related to the continuous-domain variable $m$ as
\be
g=\frac{\sqrt{\gamma\gamma'}}{6\sqrt{z}}m^3
\ee
and replace the unknown function $u_m$ with the quantity 
\be
w(g)=u_m/g^{5/6}
.
\ee 
Upon doing so, Eq.\eqref{eq:s3.5} is transformed to the Bessel equation for $w(g)$ that has a general solution of the form 
\be
w(g)=C_1 I_{5/6}(g)
+ C_2I_{-5/6}(g),
\ee 
where $I_{\alpha}(g)$ is the $\alpha$th-order modified Bessel function of the first kind. This result allows us to write the general solution for $u_m$ as
\begin{equation}\label{eq:s4} 
u(g)=g^{5/6}[C_1 I_{5/6}(g)
+ C_2I_{-5/6}(g)]
,
\end{equation}
where $C_1$ and $C_2$ are constants that must be determined from the boundary conditions. 

From Eq.\eqref{eq:s3}, we know that constant prefactor of $u$ does not affect $\Gamma(m)$ values. 
Therefore, the only quantity that remains to be determined is the ratio $C_2/C_1$. 
When $m$ varies from $0$ to $m_{\ast}= (\gamma/\gamma')^{1/4}$, $g$ varies from $0$ to 
\be
g_{\ast} = \gamma^{5/4}/(6\gamma'^{1/4}\sqrt{z}).
\ee 
The boundary condition for $u(g)$ at the right end of the interval is given by Eq.\eqref{eq:s1.5}, that is 
\begin{equation}\label{eq:s4.5}
  -\frac{m_{\ast}\gamma}{6g_{\ast}}\left.\frac1{u} \frac{du}{dg}\right|_{g=g_{\ast}} =\Gamma(m_{\ast})>0
\end{equation}
When $k \ll \frac{\gamma^{5/4}}{3v_F \gamma'^{1/4}}\equiv \kappa_{>}$, Eq.\eqref{eq:s4.5} requires $\Gamma > 0$ for $g_{\ast} \gg 1$.
However, the large-argument asymptotic expansions for $I_\alpha(g)$ are exponentially growing in exactly the same way for $\alpha$ and $-\alpha$.
This results in a monotonically increasing $u$, 
producing unphysical values $\Gamma(m_\star)<0$.
The only way to resolve this conflict is to pick $C_2\simeq -C_1$,
then all terms in the asymptotic expansion cancel out.


We therefore arrive at Eq.\eqref{eq:s4} with $C_1=-C_2$. 
The quantity 
of interest is $\Gamma(k)=\Gamma_{m=2}$ corresponds to the solution $u(g)$ of the  differential equation, Eq.\eqref{eq:s3.5}, at the variable $g$ value 
\be
g|_{m=2} = \frac{4}{3} \frac{\sqrt{\gamma\gamma'}}{\sqrt{z}}
.
\ee
Importantly, since for $k \gg  
\kappa_{<}=\frac{\sqrt{\gamma\gamma'}}{v_F}$, this quantity is small, $g|_{m=2} \ll 1$. Therefore, we can expand $u(g)$ in $g\ll 1$. This can be done using the Taylor series 
\be
I_\nu(g)=
(g/2)^\nu\sum_{m=0}^\infty\frac{
(g^2/4)^m}{m!\Gamma(\nu+m+1)},
\ee
where $\Gamma(x)$ is the gamma function \cite{Gamma_vs_Gamma}. This yields
\begin{equation}\label{eq:u_asymptotic}
  u(g)=-\frac{2^{\frac{5}{6}}}{\Gamma(\frac{1}{6})}+\frac{2^{\frac{5}{6}}}{\Gamma(\frac{11}{6})}\lp\frac{g}{2}\rp^{5/3}-\frac{1}{2^{\frac{1}{6}}\Gamma(\frac{7}{6})}\lp\frac{g}{2}\rp^{2}+\cdots
.
\end{equation}
%
Plugging this $u(g)$ in Eq.\eqref{eq:s3} 
gives the dependence 
\begin{equation}\label{eq:s6}
  \Gamma(k) = 
  \frac{\Gamma(\frac{1}{6})}{3^{\frac{2}{3}}2^{\frac{4}{3}}\Gamma(\frac{5}{6})}\frac{\gamma}{(\gamma\gamma')^{\frac{1}{6}}}
  z^{\frac16} 
, 
\end{equation}
valid up to a $T$-independent correction describing ballistic conductivity at $T=0$.
The $T$ dependence of the even and odd rates, $\gamma\sim T^2$ and $\gamma'\sim T^4$, yields a $T$-linear dependence, in agreement with the numerical results presented in the main text. 

We can now check consistency of our continuous-domain approach. The validity condition introduced above, 
$|u^\prime/u| \ll 1$, applied to the dependence 
in Eq. \eqref{eq:u_asymptotic}, holds when 
$\gamma^\prime \gamma \ll z$. This inequality coincides with the criterion for the super-FL regime existence, as expected.


\begin{thebibliography}{1}
\bibitem{Gamma_vs_Gamma} The Gamma functions in Eqs.\eqref{eq:u_asymptotic} and \eqref{eq:s6} are not to be confused with the notation $\Gamma_m(k)$ for continued fractions.
\end{thebibliography}
\end{document}